\documentclass[sigconf]{acmart}

\usepackage{balance}
\usepackage{enumitem}
\usepackage{booktabs} 
\usepackage{macros}
\usepackage{tabularx}
\usepackage{amsmath}
\usepackage{times}

\usepackage{color}

\setcopyright{none}

\acmDOI{10.475/123_4}

\acmISBN{123-4567-24-567/08/06}

\acmConference[FATREC 2018]{ACM}{October 2018}{Vancouver, Canada}
\acmYear{2018}
\copyrightyear{2018}

\acmPrice{15.00}

\begin{document}

\title{A Fairness-aware Hybrid Recommender System}

\author{Golnoosh Farnadi}
\affiliation{UC Santa Cruz}
\email{gfarnadi@ucsc.edu}

\author{Pigi Kouki}
\affiliation{UC Santa Cruz}
\email{pkouki@soe.ucsc.edu}

\author{Spencer K. Thompson}
\affiliation{UC Santa Cruz}
\email{spkthomp@ucsc.edu}

\author{Sriram Srinivasan}
\affiliation{UC Santa Cruz}
\email{ssriniv9@ucsc.edu}

\author{Lise Getoor}
\affiliation{UC Santa Cruz}
\email{getoor@soe.ucsc.edu}

\renewcommand{\shortauthors}{G. Farnadi et al.}

\begin{abstract}
Recommender systems are used in variety of domains affecting people's lives. This has raised concerns about possible biases and discrimination that such systems might exacerbate. There are two primary kinds of biases inherent in recommender systems: observation bias and bias stemming from imbalanced data. Observation bias exists due to a feedback loop which causes the model to learn to only predict recommendations similar to previous ones. Imbalance in data occurs when systematic societal, historical, or other ambient bias is present in the data.
In this paper, we address both biases by proposing a hybrid fairness-aware recommender system.
Our model provides efficient and accurate recommendations by incorporating multiple user-user and item-item similarity measures, content, and demographic information, while addressing recommendation biases. 
We implement our model using a powerful and expressive probabilistic programming language called \textit{probabilistic soft logic}. 
We experimentally evaluate our approach on a popular movie recommendation dataset, showing  that our proposed model can provide more accurate and fairer recommendations, compared to a state-of-the art fair recommender system. 

\end{abstract}


\maketitle

\section{Introduction}
\label{Introduction}

Targeted recommendations have become increasingly important to business owners in order to reach their potential customers. Such systems are used in a variety of domains such as commerce, employment, dating, health, education, and governance. However, when targeting users, biases can have a negative impact on subgroups of users. For instance, one study~\cite{datta2015automated} shows that female users of Google have a lower chance of accessing hiring ads for high-paying executive jobs.  
The study of bias and fairness in machine learning is an emerging research area that is receiving increasing attention \cite{bozdagEIT2013, dworkTCSC2012,pedreshiKDD2008}.  
Methods mitigating unfairness in machine learning systems~\cite{kamiranICDM2010,kamishimaKDD2012,zemelICML2013}
can be extended to the case of fairness-aware recommender systems.


There are a variety of definitions of fairness~\cite{hardt2016equality,Pedreschi:2012}. Defining fairness, especially for recommender systems, is challenging. In this paper, 
we assume that someone has given us the definitions of which attributes/sub-population we want to maintain fairness towards, and present a scalable, declarative formulation for achieving fairness relative to the given subgroups. In the fairness domain, a population of vulnerable individuals known as the \emph{protected group}, and can be defined by an attribute value upon which discrimination is based (such as gender, ethnicity, or religion).
A fair recommender system should provide rankings to the protected group that are the same as the \emph{unprotected group}. The majority of popular recommender system algorithms (e.g., collaborative-filtering) make use of user behavior to generate recommendations. 
Powerful as they are, these methods usually inherit the biases that exists in the data which may cause the system to present unfair recommendations. 

There are two primary kinds of bias that can be inherited from data: observation bias and bias that comes from imbalance in data \cite{siruiNIPS2017}. 
Observation bias is due to the existence of a feedback loop in the system.  An item displayed by the recommender system may result in an action, which is then used to retrain the model. This reinforces the recommender system's ranking algorithm to show more items similar to previous recommendations. If a user is never exposed to an item, the user cannot provide an opinion on it. For example, if a user on a movie streaming website was never shown a movie from the action genre, it is difficult for the system to know the user's interest level in action movies.
In contrast, imbalance in data is caused when a systematic bias is present due to societal, historical, or other ambient biases. Since the model is unaware of such biases, addressing them is not straightforward. For example, in job recommendation, due to social bias, the data contain a lot of evidence indicating that nursing is a successful profession for females. However, this does not mean that female users must be recommended to
receive recommendations for nursing positions 
in order to be successful. 

These biases have been explored and addressed in different contexts with use of multi-arm bandits and diversity-based recommendations~\cite{wang2017online,wasilewski2016incorporating,Delft2017PresentingDA}. 
Although these approaches tend to handle bias by increasing the diversity of a recommender system, they do not directly address the issue of fairness.  More recently, fairness in recommender systems has been explored through the use of fairness metrics. 
For instance, Yao and Huang \cite{siruiNIPS2017}, show that fairness can be both measured and imposed on a matrix factorization (MF) method, using five different fairness metrics. Burke et al. \cite{Burke2017BalancedNF} recently introduced an approach that aspires for both personalization and fairness via neighborhood balancing with a sparse linear method. 
Other work studies the issue of modeling in the presence of gender-imbalanced data. 
As an example, Sapiezynski et al. \cite{Sapiezynski2017AcademicPP} found that gender representation-imbalance in academic data on students led to a higher accuracy in detecting struggling male students, as opposed to their female classmates.  

In this work, we first start by using a hybrid recommender system, called HyPER \cite{kouki:2015} to produce recommenations. HyPER incorporates a variety of signals including user-user similarities, item-item similarities, and content and demographic  information. We then extend HyPER to a fairness-aware recommender system by addressing observation bias and biases coming from imbalanced data.  
We implement our fairness-aware recommender system as a single unified model, by using a probabilistic programming language called probabilistic soft logic (PSL) \cite{bach:jmlr17}. PSL has been used for providing hybrid recommendations  \cite{kouki:2015}, for providing a fairness-aware framework in relational settings \cite{golnoosh:ai18}, 
and a variety of other tasks. In this work, we unify these two lines of work and propose a fairness-aware hybrid recommendation system. We make use of a constraint-based approach to fairness which extends PSL with a new maximum a posteriori (MAP) inference algorithm that 
maximizes the a posteriori values of unknown variables subject to fairness guarantees using a set of hard fairness constraints. 
Like previous work, we model various protected and unprotected groups with relational dependencies in the model; however, in our proposed work, we address biases in recommender systems with a set of soft fairness constraints that are directly expressed in the model.
We design two sets of fairness constraints with latent variables that are able to: 1) detect and address biases in item ratings coming from imbalanced data and 2) integrate rules that address biases coming from item group ratings to prevent observation bias. These two sets of constraints are able to capture relational dependencies among users and items to collectively predict accurate ratings for both protected and unprotected groups.

In this paper, we make the following contributions: 
1) we present a probabilistic programming approach for building fair hybrid recommender systems;
2) we experimentally study fairness on the popular MovieLens dataset; 3) we
show that a fair recommender system can outperform a recommender system not trained for fairness on both accuracy and fairness evaluation metrics; 
and 4) we experimentally show that our fair recommender system surpasses the current state-of-the-art fair recommender system in both accuracy and a variety of fairness metrics. 

The remainder of the paper is structured as follows: 
In Section \ref{Model}, we present our model in detail. We start with an overview of the modeling language that we use to build a fair movie recommender system, i.e., PSL (Section \ref{PSL}). We briefly describe the movie recommender system that we use, i.e., HYPER (Section \ref{PSLModel}), and then we explain how we can make it fair (Section \ref{FairPSLModel}). In Section \ref{Evaluation} we present our evaluation results. Finally, we conclude with a discussion and our plans for future work in Section \ref{Conclusions}.

\section{Approach}
\label{Model}

In this section, we describe how we extend an existing hybrid recommender system to provide fair recommendations. 
We first introduce the modeling framework that we use to define our model, called probabilistic soft logic (PSL). 
PSL is a declarative language that uses first-order logic to define a model. 
We choose PSL to propose a fairness-aware recommender system because its expressiveness allows us to model recommender systems as well as fairness constraints in a unified model.
Next, we describe how we define a hybrid movie recommender system using the basic principles of an existing hybrid recommender system (HyPER). 
Finally, we discuss how we extend our recommender system to account for fairness by using a set of PSL rules capturing fairness with relational dependencies between users and items. 

\subsection{PSL}
\label{PSL}

Probabilistic soft logic (PSL)~\cite{bach:jmlr17} is a probabilistic programming language that uses a first-order logical rules to define a graphical model. PSL uses continuous random variables in the $[0,1]$ unit interval and specifies factors using convex functions, allowing tractable and efficient inference. 
PSL defines a Markov random field associated with a conditional probability density function over random variables $\mathbf{Y}$ conditioned on evidence $\mathbf{X}$,
\begin{align}
P(\mathbf{Y}|\mathbf{X}) \propto \exp\Big( -\sum_{j=1}^{m}w_j\phi_j(\mathbf{Y},\mathbf{X})\Big)  \mbox{ ,}
\label{eq:hl-mrf}
\end{align}
where $\phi_j$ is a convex potential function and $w_j$ is an associated weight which determines the importance of $\phi_j$ in the model. The potential $\phi_j$ takes the form of a \emph{hinge-loss}:
\begin{align}
\phi_j(\mathbf{Y},\mathbf{X}) = (max\{0,\ell_j(\mathbf{X},\mathbf{Y})\})^{p_j} \mbox{ .}
\label{phi}
\end{align}
Here, $\ell_j$ is a linear function of \textbf{X} and \textbf{Y}, and $p_j \in \{1,2\}$ optionally squares the potential, resulting in a \emph{squared-loss}. The resulting probability distribution is log-concave in $\mathbf{Y}$, so we can solve MAP inference via convex optimization to find the optimal $\mathbf{Y}$. 
The convex formulation of PSL is the key to efficient, scalable inference in models with many complex inter-dependencies. 

PSL derives this objective function by translating logical rules that specify dependencies between variables and evidence into hinge-loss functions. PSL achieves this translation by using the \textit{Lukasiewicz} norm and co-norm to provide a relaxation of Boolean logical connectives~\cite{bach:jmlr17}. For example, $a \Rightarrow b$ corresponds to the hinge function $\max(a - b, 0)$, and $a \wedge b$ corresponds to $\max(a + b - 1, 0)$.  
We refer the reader to~\cite{bach:jmlr17} for a detailed description of PSL.  

To illustrate PSL in the movie recommendation context, the following rule encodes that users tend to rate movies of their preferred genres highly:
\begin{equation*}
\textsc{LikesGenre}(\texttt{u},\texttt{g}) \wedge \textsc{IsGenre}(\texttt{m},\texttt{g}) \Rightarrow \textsc{Rating}(\texttt{u},\texttt{m}) \mbox{ ,}
\label{rule}
\end{equation*}
where $\textsc{LikesGenre}(\texttt{u},\texttt{g})$ is a binary observed predicate,
$\textsc{IsGenre}(\texttt{m},\texttt{g})$ is a continuous observed predicate in the interval $[0,1]$ capturing the affinity of the movie to the genre,
and $\textsc{Rating}(\texttt{u},\texttt{m})$ is a continuous variable to be inferred, which encodes the star rating as a number between 0 and 1, with higher values corresponding to higher star ratings.  For example, we could instantiate $\texttt{u}=\mbox{\texttt{Jim}}$, $\texttt{g}=\mbox{\texttt{classics}}$ and $\texttt{m}=\mbox{\texttt{Casablanca}}$.  This instantiation results in a hinge-loss potential function in the HL-MRF,
\begin{align*}
\max(& \textsc{LikesGenre}(\mbox{\texttt{Jim}},\mbox{\texttt{classics}}) \\
&+ \textsc{IsGenre}(\mbox{\texttt{Casablanca}},\mbox{\texttt{classics}})\\ 
&- \textsc{Rating}(\mbox{\texttt{Jim}},\mbox{\texttt{Casablanca}})
- 1,0) \mbox{ .}
\end{align*}
PSL has been successfully applied in various domains, such as explanations in recommender systems \cite{kouki:recsys17}, user modeling in social media \cite{farnadi2017soft}, stance prediction in online forums \cite{sridhar:acl15}, energy disaggregation \cite{tomkins:ijcai17} and knowledge graph identification \cite{pujara:iswc13}.

\subsection{PSL Recommendations Model}
\label{PSLModel}

In recent work, Kouki et al. \cite{kouki:2015} introduced \hyper, a hybrid recommender system that uses PSL. 
The model consists of rules that can incorporate a wide range of information sources, such as user-user and item-item similarity measures, content information, and user and item average predictors.
\hyper~ uses the rules together with the input data to perform inference and define a probability distribution over the recommended items, capturing the extent to which a given user will like a given item.
\hyper~provides a generic and extensible recommendation framework with the ability to incorporate other sources of information that may be available in different domains. In this work, we focus on movie recommendations. We use a subset of all the rules proposed in HyPER, and we add rules to leverage dataset-specific information available in our movie dataset.
We propose a hybrid movie-recommender system which consists of the following rules:

\subsubsection{Mean-Centering Priors}
\label{sec:priors}
We first encode rules in our model that encourage the ratings to be close to the average for each user and each item.
Each individual user of a recommender system has his/her own preferences in rating items. Likewise, each item's rating is influenced by its overall popularity.
To capture this information, we introduce the following rules:
\begin{equation*}
{\small
\begin{split}
\textsc{AverageUserRating}(\texttt{u}) &\Rightarrow ~~\textsc{Rating}(\texttt{u},\texttt{i})\\
\neg\textsc{AverageUserRating}(\texttt{u}) &\Rightarrow \neg \textsc{Rating}(\texttt{u},\texttt{i})
~\\
\textsc{AverageItemRating}(\texttt{i}) &\Rightarrow ~~\textsc{Rating}(\texttt{u},\texttt{i})\\
\neg\textsc{AverageItemRating}(\texttt{i}) &\Rightarrow \neg \textsc{Rating}(\texttt{u},\texttt{i}) \mbox{ .}
\end{split}
}
\end{equation*}
The predicate $\textsc{Rating}(\texttt{u},\texttt{i})$ takes a value in the interval $[0,1]$ and represents the normalized value of the rating that a user $\texttt{u}$ gave to an item $\texttt{i}$. 
The predicate $\textsc{AverageUserRating}(\texttt{u})$ represents the average of the ratings over the set of items that user $\texttt{u}$ provided in the training set. Similarly, $\textsc{AverageUserRating}(\texttt{i})$ represents the average of the user ratings an item $\texttt{i}$ has received.  The pair of PSL rules per-user and per-item  penalizes the predicted rating for being different from this average.

\subsubsection{Neighborhood-based Collaborative Filtering}
\label{sec:userbased}
We define PSL rules that capture the basic principle of the neighborhood-based approach. We introduce the following user-based collaborative filtering rule to capture the intuition that similar users give similar ratings to the same items:
\begin{equation*}
{\small
\begin{split}
\textsc{SimilarUsers}_\textsc{sim}(\texttt{u}_1,\texttt{u}_2) \wedge \textsc{Rating}(\texttt{u}_1,\texttt{i}) \Rightarrow& ~~\textsc{Rating}(\texttt{u}_2,\texttt{i})\mbox{ .}
\end{split}
}
\label{sim_users}
\end{equation*}

There are several ways one can do this in PSL, here we the predicate $\textsc{SimilarUsers}_\textsc{sim}(\texttt{u}_1,\texttt{u}_2)$ is binary, with value 1 iff $\texttt{u}_1$ is one of the $k$-nearest neighbors of $\texttt{u}_2$. 
The above rule represents a template for hinge functions which reduces as the probability of predicted ratings as the difference between $\textsc{Rating}(\texttt{u}_2,\texttt{i})$ and $\textsc{Rating}(\texttt{u}_1,\texttt{i})$ increases, for users that are neighbors.
Similarly, we can define PSL rules to capture the intuition of item-based collaborative filtering methods, namely that similar items should have similar ratings from the same users:
\begin{equation*}
{\small
\begin{split}
\textsc{SimilarItems}_\textsc{sim}(\texttt{i}_1,\texttt{i}_2) \wedge \textsc{Rating}(\texttt{u},\texttt{i}_1) & \Rightarrow ~~\textsc{Rating}(\texttt{u},\texttt{i}_2)\mbox{ .}
\end{split}
}
\label{sim_items}
\end{equation*}
As before, the predicate $\textsc{SimilarItems}_\textsc{sim}(\texttt{i}_1,\texttt{i}_2)$ is binary, with value 1 iff $\texttt{i}_1$ is one of the $k$-nearest neighbors of $\texttt{i}_2$. 
The similarities can be calculated with any similarity measure $\textsc{sim}$.
In this model, we use the most popular similarity measures in the neighborhood-based recommendations literature~\cite{ning:15}.
More specifically, we apply cosine similarity measures to calculate similarities between users and items; for the items we additionally apply the adjusted cosine similarity using a
Pearson's correlation measure.

\subsubsection{Using Additional Sources of Information}
\label{sec:additional}
The movie dataset that we use offers demographic information about the users and content information about the items. We can use this information to define similar users using demographic information and similar items using content. We introduce the following rules:
\begin{equation*}
{\small
\begin{split}
\textsc{SimilarUsers}_\textsc{Demo}(\texttt{u}_1,\texttt{u}_2) \wedge \textsc{Rating}(\texttt{u}_1,\texttt{i}) \Rightarrow& ~~\textsc{Rating}(\texttt{u}_2,\texttt{i})
~\\
\textsc{SimilarItems}_\textsc{Content}(\texttt{i}_1,\texttt{i}_2) \wedge \textsc{Rating}(\texttt{u},\texttt{i}_1) \Rightarrow& ~~\textsc{Rating}(\texttt{u},\texttt{i}_2)\mbox{ .}
\end{split}
}
\label{side_info}
\end{equation*}
In the first rule, the predicate $\textsc{SimilarUsers}_\textsc{Demo}(\texttt{i}_1,\texttt{i}_2)$ represents users that have similar demographic  features (which in our case is age, gender, and occupation). 
In the second rule, the predicate $\textsc{SimilarItems}_\textsc{Content}(\texttt{i}_1,\texttt{i}_2)$ represents items that have similar content-based features (which in our case in the genre of a movie). 

\subsubsection{Negative Prior}
\label{sec:prior}
In recommender systems, there is usually a huge number of items available (e.g., movies). However, every single user has rated a small number of these items. 
To model our general belief that it is unlikely for a user to rate a movie, we introduce the following prior rule:
\begin{equation*}
{\small
\neg \textsc{Rating}(\texttt{u},\texttt{i}) \mbox{ .}
}
\end{equation*}

\subsection{Fair PSL Recommendations Model}
\label{FairPSLModel}

The power of our PSL recommender system lies in the fact that fairness can also be modeled with a set of logical rules. This is particularly important given the diverse set of applications powered by recommender systems and that fairness in many of those applications is multi-faceted~\cite{burke17aa}. For example, a recommender system suggesting job applications needs to ensure that two applicants with a similar professional profile receive similar job recommendations. 
At the same time, the recommender system needs to ensure market diversity and avoid monopoly domination by giving similar chance to new companies to get a reasonable share of recommendations even though they have had fewer job offerings
compared to the established companies. In this work, we take into account the disparate impact of recommendation on protected classes of recommendation users.

To encode fairness in our model, we first introduce protected and unprotected groups. The predicate $\textsc{Protected}(\texttt{u})$ is binary, which indicates whether a user belongs to the protected group with value $1$, or unprotected group with value $0$.
 Note  though that the protected group could be any attribute and it can be either an observed attribute in the data or a latent attribute. Here, we consider all female users to be our protected group and all male users to be our unprotected group. 
The goal of a fair recommender system is to provide fair ratings depending on the fairness metric used, for both protected and unprotected groups.

\subsubsection{Imbalance in Data}

In recommender systems, various types of users may have a tendency to only rate particular items. For instance, female users may be more likely to shop for clothes, while male users may buy tools with higher frequency. 
If a recommender system has access only to an imbalanced dataset, it may never recommend a particular item to a specific group of users. To avoid such bias in our model, we define the following rules:

\begin{equation*}
{\small
\begin{split}
\textsc{Protected}(\texttt{u}) \wedge \textsc{Rating}(\texttt{u},\texttt{i}) \Rightarrow& \textsc{ProtectedItemRating}(\texttt{i})
~\\
\neg \textsc{Protected}(\texttt{u}) \wedge \textsc{Rating}(\texttt{u},\texttt{i}) \Rightarrow& \textsc{UnprotectedItemRating}(\texttt{i})\mbox{ .}
\end{split}
}
\label{side_info}
\end{equation*}

At a high level, we introduce two latent variables for each item, i.e.,  $\textsc{ProtectedItemRating}(\texttt{i})$ and $\textsc{UnprotectedItemRating}(\texttt{i})$. These two predicates capture ratings from protected and unprotected users for each item in the data. To encode fair ratings for both groups, we add the following constraints to the model which force the value of these two latent variables for each item to be equivalent, for both protected and unprotected groups:
\begin{equation*}
{\small
\begin{split}
 \textsc{ProtectedItemRating}(\texttt{i}) \Rightarrow& \textsc{UnprotectedItemRating}(\texttt{i})
~\\
 \textsc{UnprotectedItemRating}(\texttt{i}) \Rightarrow& \textsc{ProtectedItemRating}(\texttt{i})\mbox{ .}
\end{split}
}
\label{side_info}
\end{equation*}

Using the above rules, we are able to balance the ratings for both types of users by un-biasing the ratings for each item. Extending the recommender model that we described in Section~\ref{PSLModel} with these rules enables us to address imbalanced data biases.
For the case of movie recommendation, our protected and unprotected groups are female users and male users respectively. Therefore, we can replace predicate $\textsc{Protected}(\texttt{u})$ with $\textsc{IsFemale}(\texttt{u})$ that indicates whether a user is female or male.

\subsubsection{Observation Bias}

In addition to bias coming from imbalance in the data, users may prefer items belong to a certain item group. For example, for the item group ``genre'' in the context of movie recommendations, female users may be more likely to rate romance movies, while male users may rate action movies with higher frequency. If users have never recommended a particular item, they will likely never provide rating data for that item. To avoid such observation bias we introduce fairness rules for item groups in the model. 
Similar to the fairness rules of the previous section, 
we introduce latent variables for each item group with the following rules: 
\begin{equation*}
{\small
\begin{split}
\textsc{Protected}(\texttt{u}) \wedge \textsc{Rating}(\texttt{u},\texttt{i}) \wedge \textsc{ItemGroup}(\texttt{i},\texttt{g}) \\\Rightarrow \textsc{ProtectedItemGroupRating}(\texttt{g})
~\\
\neg \textsc{Protected}(\texttt{u}) \wedge \textsc{Rating}(\texttt{u},\texttt{i}) \wedge \textsc{ItemGroup}(\texttt{i},\texttt{g}) \\\Rightarrow \textsc{UnprotectedItemGroupRating}(\texttt{g})\mbox{ .}
\end{split}
}
\label{side_info}
\end{equation*}

Here, we introduce two latent variables for each item group by introducing two predicates, i.e.,  $\textsc{ProtectedItemGroup}$
$\textsc{Rating}(\texttt{g})$ and $\textsc{UnprotectedItemGroupRating}(\texttt{g})$, that capture ratings from protected and unprotected users for items within each item group. Similarly, we force the value of these latent variables for each item group in the model to be equal using the following rules:
\begin{equation*}
{\small
\begin{split}
 \textsc{ProtectedItemGroupRating}(\texttt{g}) \\\Rightarrow \textsc{UnprotectedItemGroupRating}(\texttt{g})
~\\
 \textsc{UnprotectedItemGroupRating}(\texttt{g}) \\\Rightarrow \textsc{ProtectedItemGroupRating}(\texttt{g})\mbox{ .}
\end{split}
}
\label{side_info}
\end{equation*}

Our fair recommender model includes both types of fairness rules to address observational bias, and avoid bias coming from imbalanced ratings for both protected and unprotected groups. 
The definition of an item group depends on the specific context of the recommender system. 
In the movie recommendation setting, we can use a movie's genre to define item groups. Therefore, in the above rules, we can replace the predicate $\textsc{ItemGroup}(\texttt{i},\texttt{g})$ with the predicate $\textsc{IsGenre}(\texttt{i},\texttt{g})$ to capture the affinity of a movie to a genre. Note that movie could have more than one genre, for instance the movie $\emph{Casablanca}$ has three genres: $\emph{classics}$, $\emph{drama}$, and $\emph{romance}$.

We use the rules presented in this section with the rules presented in Section~\ref{PSLModel} to collectively infer ratings for all users. Next, we present our experimental setup for evaluating our proposed fair movie recommender system.

\section{Experimental Validation}
\label{Evaluation}

\begin{table*}[!h]
    \centering
   \resizebox{\textwidth}{!}{ 
   \begin{tabular}{l || c|c || c | c | c | c | c | c}
    \hline
        {\bf Model} & {\bf RMSE}& {\bf MAE} & {\bf Overesti-} & {\bf Absolute} & {\bf Non-Parity} & {\bf Underesti-} & {\bf Value} & {\bf Balance} \\
        & \small{(SD)} &\small{(SD)}  & {\bf mation} \small{(SD)} & \small{(SD)}& \small{(SD)}& {\bf mation} \small{(SD)} &\small{(SD)} &\small{(SD)}\\
      \hline
      (MC) Baseline & 0.997 \small{(0.003)} & 0.794 \small{(0.002)}& 0.280 \small{(0.001)}& 0.302 \small{(0.002)}& 0.144 \small{(0.001)} & {\bf 0.104} \small{(0.001)}&  0.385 \small{(0.002)} & 0.192 \small{(0.002)}  \\
      \hline
        MF~\cite{siruiNIPS2017} & 0.944 \small{(0.002)} & 0.760 \small{(0.002)} & 0.256 \small{(0.001)} & 0.282 \small{(0.001)} & \colorbox{white}{{ 0.084} \small{(0.000)} }& 0.139 \small{(0.001)}  & 0.395 \small{(0.002)} & 0.198 \small{(0.002)} \\
        Fair MF (non-parity)~\cite{siruiNIPS2017} & 0.945 \small{(0.002)} & 0.760 \small{(0.002)} & 0.252 \small{(0.001)} & 0.281 \small{(0.001)} & {\bf 0.083} \small{(0.000)} & 0.145 \small{(0.001)}  & 0.396 \small{(0.002)}  & 0.199 \small{(0.001)}  \\
        Fair MF (value)~\cite{siruiNIPS2017}& 0.948 \small{(0.002)} & 0.762 \small{(0.002)} & 0.250 \small{(0.002)} & 0.279 \small{(0.002)} & \colorbox{white}{{0.131} \small{(0.000)}} & 0.140 \small{(0.000)}  & 0.390 \small{(0.002)} & 0.197 \small{(0.001)}\\
         \hline
         (MC+CF) PSL & 0.922 \small{(0.002)} & 0.734 \small{(0.001)} & {\bf 0.144} \small{(0.001)}& 0.261 \small{(0.001)} & 0.224 \small{(0.000)}& 0.210 \small{(0.001)}& {\bf 0.354} \small{(0.002)} & {\bf0.177} \small{(0.001)}\\
        (MC+CF+DC) PSL~\cite{kouki:2015} & 0.916 \small{(0.002)} & 0.732 \small{(0.002)} & {0.180} \small{(0.001)}& 0.260 \small{(0.002)} & 0.206 \small{(0.001)}&  0.177 \small{(0.001)}& {0.357} \small{(0.002)} & 0.179 \small{(0.001)}\\
        \hline       

        Fair (MC+CF) PSL & {\bf 0.908} \small{(0.002)} & {\bf 0.727} \small{(0.001)} & 0.158 \small{(0.001)} & {0.251} \small{(0.001)} & 0.159 \small{(0.001)} & 0.201 \small{(0.001)} & {0.358} \small{(0.002)} & 0.180 \small{(0.001)} \\
        
        Fair (MC+CF+DC) PSL & 0.911 \small{(0.002)} & 0.730 \small{(0.002)} & 0.177 \small{(0.001)} & {\bf0.250} \small{(0.001)} & 0.160 \small{(0.001)} & 0.180 \small{(0.001)} & { 0.357} \small{(0.002)} & 0.179 \small{(0.001)}\\
         \hline
    \end{tabular}
    }
    \caption{Overall Performance of different PSL and fair MF models. Numbers in parenthesis indicate standard deviations. For each metric, we report the best value in bold. For all metrics, the smaller the value, the more accurate/fair the model is.}
    \label{tab:results}
    \vspace{-0.3cm}
\end{table*}

\subsection{Dataset Description}
\label{dataset}
For our experiments, we use the MovieLens 1M dataset \cite{movielens}, which has ~1M ratings (ranging from 1 to 5) for ~4k movies, from ~6k users.  
Demographic data for the users (e.g., gender, age, occupation) and metadata on the movies (e.g., genre) are also provided.  
For movies, we follow the preprocessing steps proposed by Yao and Huang \cite{siruiNIPS2017} and consider only movies that are tagged with at least one of the following 5 genres: \textit{action, romance, crime, musical, sci-fi}. These genres have a relatively large difference between the number of ratings by males and females, as well as a noticeable difference in average rating by each gender. This yields a subset of the dataset that expresses a strong population imbalance and gives the potential for an unfair recommender system. 
This is evident from the gender-based statistics of movie genres reported in Table 2 of \cite{siruiNIPS2017}. For example, the number of ratings per female user for romantic movies is $54.67$, while for men it is $36.97$. In another example, the number of ratings per female user for sci-fi movies is $31.19$, while for male users it is $50.46$.
Again, following the filtering process proposed in Yao and Huang, we further filter the dataset by only considering users that rated more than $50$ movies.
These preprocessing steps produce a subset of the original Movielens 1M dataset, consisting of $443,079$ ratings for $1,305$ movies from $2,965$ users. 


\subsection{Evaluation Metrics}
\label{metrics}
To measure the accuracy of the movie recommender system, we report the root mean squared
error (RMSE) and the mean absolute error (MAE).
To measure the fairness (or unfairness) of the movie recommender system, we use the popular demographic parity measure \cite{Calders:2009} and a set of new metrics, recently introduced by Yao and Huang \cite{siruiNIPS2017}. 
These are the fairness metrics that we report in our experimental evaluation:


\begin{itemize}[leftmargin=*,noitemsep,topsep=0pt]
\item \textit{Non-parity unfairness}: measures the absolute unfairness in making predictions for two groups (the protected and  unprotected groups). This metric is computed  as the absolute difference between the overall average ratings of users belonging to the unprotected group and those of users belonging to the protected group.
\item \textit{Value unfairness}: measures the inconsistency in signed estimation error across the protected and unprotected user groups. 
This metric becomes large when predictions for one group are consistently overestimated while predictions for the other group are consistently underestimated.
\item \textit{Absolute unfairness}: measures the inconsistency in absolute estimation error across user groups. This metric is sign-agnostic and its value becomes large if one group of users consistently receives more accurate recommendations than the other.
\item \textit{Underestimation unfairness}: measures the inconsistency in how much the predictions underestimate the true ratings. This metric becomes large when the recommender system constantly predicts lower rating values than the true ratings. 
\item \textit{Overestimation unfairness}: measures the inconsistency in how much the predictions overestimate the true ratings. This metric is the opposite of underestimation unfairness, i.e., when overestimation unfairness increases in a system then underestimation unfairness decreases (and vice versa). This metric becomes large when the recommender system constantly predicts higher rating values than the true ratings.
\item \textit{Balance unfairness}: measures the inconsistency in how much the predictions overestimate and underestimate the true ratings. This metric is the average of underestimation and overestimation unfairness. 
\end{itemize}

\subsection{Experiments}
\label{Experiments}


We evaluate the following different versions of the PSL model:
\begin{itemize}[leftmargin=*,noitemsep,topsep=0pt]
\item \textit{Mean-centering (MC) model}: This model uses the mean-centering priors for users and items described in Section ~\ref{sec:priors} and the negative prior (Section \ref{sec:prior}). Note that this model has no rules to model relational dependencies, and therefore is not using PSL capabilities. We call this model \textbf{MC Baseline}.
\item \textit{Mean-centering (MC) and collaborative filtering (CF) PSL model}: We enrich the above model by adding user and item similarity rules explained in Section \ref{sec:userbased}. We note again that these similarities are computed using only collaborative filtering information.  
The number of similar users and items is typically set to between 20 and 50 in the literature \cite{ning:15}, and here for each user we use the 20 most similar neighbors. This selection of k=20 applies to all the nearest-neighbor similarities that we use. 
We call this model \textbf{MC+CF PSL}. 
\item \textit{Mean-centering (MC), collaborative filtering (CF), and demographic/content (DC) PSL model}: 
We extend the above model by adding additional information about users and items (Section \ref{sec:additional}). 
Specifically, we use demographic information on individual users, i.e. gender, age, and occupation, to compute the cosine similarity between user pairs. 
Item similarities are computed for each pair of movies, using cosine similarity among vector representations of the movies genres. 
We call this model \textbf{MC+CF+DC PSL}.
\item \textit{Fair Mean-centering and collaborative filtering PSL model}: We enrich the model \textbf{MC+CF PSL} by adding all the fair rules described in Section \ref{FairPSLModel}. 
We call this model \textbf{Fair MC+CF PSL}.
\item \textit{Fair Mean-centering, collaborative filtering, and demographic/content PSL model}: We enrich the model \textbf{MC+CF+DC PSL} by adding all the fair rules described in Section \ref{FairPSLModel}.
We call this model \textbf{Fair MC+CF+DC PSL}.
\end{itemize}

We compare our fair PSL models with the fair state-of-the-art models proposed by Yao and Huang \cite{siruiNIPS2017}.
Yao and Huang proposed six different models, where five models optimized for different fairness metric.
We ran all six models and we report the results of the following three: 
1) \textbf{MF}, a matrix factorization algorithm that does not optimize for fairness (we include this model as the simplest MF without fairness), 
2) \textbf{Fair MF (non-parity)} which optimizes for the non-parity unfairness metric (this model performed the best in terms of RMSE and MAE in the $5$ splits of the dataset that we used for our experiments),
 and 3) \textbf{Fair MF (Value)} which optimizes for the value unfairness metric (this model performed the best in the $5$ splits of the dataset that Yao and Huang \cite{siruiNIPS2017} operated on). 
To run these models, we used the default values for regularization and epochs, i.e., reg=$0.001$ and epochs=$100$.

We compute the metrics described in Section \ref{metrics} by performing five-fold cross-validation in the filtered MovieLens dataset described in Section ~\ref{dataset}. We report the average cross-validated errors and unfairness values along with the standard deviation for all the models described above using the same split. We report our results in Table \ref{tab:results}.
For each metric, we report the best value in bold. For all metrics, the smaller the value, the more accurate/fairer the model is.

\subsection{Results}
\label{Discussion}
We observed the following from Table \ref{tab:results}:

\textbf{PSL shows improved accuracy compared to MF methods}: A first clear conclusion from the results is that the all PSL models outperform all three MF (fair and non-fair) models on accuracy metrics. 
With one exception (the simplest PSL model, {MC Baseline} that only uses average ratings), PSL produces a statistically significant improvement in
both RMSE and MAE as measured by a paired t-test with $\alpha=0.05$.

\textbf{Adding fairness rules improves the performance of the PSL models}:
The addition of fairness rules to the model 
{MC+CF PSL}  (which results in the model {Fair MC+CF PSL}) results in a relatively small decrease of $0.014$ (absolute value) for the RMSE and $0.007$ (absolute value) for the MAE. 
Similarly, the addition of fairness rules to the model {MC+CF+DC PSL}  (which results in the model {Fair MC+CF+DC PSL}) results in a decrease of $0.005$ (absolute value) for the RMSE and $0.002$ (absolute value) for the MAE. 
We do not observe the same behavior for the different fair and non-fair matrix factorization models. For these cases, when trying to optimize for different fairness metrics, we observe a very small increase for the RMSE and MAE.
In particular, when trying to optimize the simple {MF} model for non-parity unfairness(which results in the model {Fair MF (non-parity)}) we observe a small increase of $0.001$ for the RMSE, while the MAE stays the same. 
Similarly, when trying to optimize the simple {MF} model for value unfairness metric (which results in the model {Fair MF (value)}) we observe a small increase of $0.004$ for the RMSE and a small increase of $0.002$ for the MAE. 

\textbf{Fair PSL models outperform fair MF models w.r.t.  balance unfairness}: PSL models (except for the (MC) Baseline) are better in avoiding underestimating the true ratings of the female users, while MF models are better in avoiding overestimating the female ratings. However, by looking at the balance unfairness metric, PSL models produce more balanced ratings for female and male users when compared to MF methods.

\textbf{Fair MF models outperform fair PSL models w.r.t.  non-parity unfairness}: All MF methods perform significantly better for the non-parity unfairness metric when compared to the PSL models. Also, optimizing for non-parity unfairness in MF causes an increase or no change in almost all the other unfairness metrics, which is consistent with the results presented in~\cite{siruiNIPS2017}.
For PSL, we note that fair PSL models perform significantly better than non-fair PSL models with respect to non-parity unfairness metric.



\textbf{There is no model that can be fair in all metrics}: There is always a trade-off between various fairness measures. Each recommender system, according to its goal, can choose a setting which satisfies its needs. According to the results presented in Table~\ref{tab:results}, there is no method that outperforms all fairness metrics. However, all PSL models (except for the (MC) Baseline model) outperform MF models in both performance and fair predication for the following metrics: RMSE, MAE, absolute unfairness, value unfairness, and balance unfairness. Also, our proposed fair PSL models outperform on all models in decreasing RMSE and MAE when predicting the true ratings of all users, and, at the same time, they consistently predict accurate ratings for female and male users. 




\section{Conclusions}
\label{Conclusions}

In this paper, we proposed a fairness-aware hybrid recommender system that integrates multiple sources of user and item data to accurately recommend items to users, and  addresses observation bias and biases coming from imbalance in the data. We implemented our system in a unified model with an expressive language, called \textit{probabilistic soft logic}. Empirical evaluation on the movie recommendation domain shows that our proposed model is able to offer more accurate and, oftentimes, fairer recommendations  compared to a state-of-the-art fair recommender system.

There are many avenues for expanding our work.
In addition to the fairness rules that we proposed in our model, we plan to extend our fairness-aware recommender system with other rules to address other types of bias, such as biases of item providers or explicit bias by advertisers. Bias by advertisers has the potential to have a polarizing affect on recommendations. In certain cases, these biases may not stem from imbalanced data but rather from the marketing practices used. 
Moreover, in a real-world recommender system setting, the user-item matrix is very sparse, contrary to the sample of the MovieLens dataset that we operated on. 
We plan to explore the robustness of our approach when data sparsity is present.
Finally, we are interested in applying our solution to other domains where fairness has  legal and policy implications, such as the job recommendation setting.

\section*{Acknowledgements}
This material is based upon work supported by the National Science Foundation under Grant Numbers CCF-1740850 and IIS-1703331, and by the US Army Corps of Engineers Research and Development Center under Contract Number W912HZ-17-P-0101.


\end{document}